\documentclass{PoS}
\usepackage{amsmath}
\usepackage{caption}

\usepackage{tikz}
\usetikzlibrary{positioning,arrows}
\usetikzlibrary{decorations.pathmorphing}
\usetikzlibrary{decorations.markings}
\usetikzlibrary{shapes.geometric}
\usepackage{endnotes}
\tikzset{
    vector/.style={decorate, decoration={snake}, draw},
    provector/.style={decorate, decoration={snake,amplitude=2.5pt}, draw},
    antivector/.style={decorate, decoration={snake,amplitude=-2.5pt}, draw},
    fermion/.style={draw=black,
      postaction={decorate},decoration={markings,mark=at position .55
        with {\arrow[draw=black]{>}}}}, 
    fermionbar/.style={draw=black, postaction={decorate},
                       decoration={markings,mark=at position .55 with {\arrow[draw=black]{<}}}},
    fermionnoarrow/.style={draw=black},
    gluon/.style={decorate, draw=black,decoration={coil,amplitude=2pt, segment length=2.5pt}},
    zboson/.style={decorate, draw=gray,decoration={snake,amplitude=2pt, segment length=3pt}},    
    photon/.style={decorate, draw=black,decoration={snake,amplitude=2pt, segment length=5pt}},
    higgs/.style={dashed,draw=black}, 
    scalar/.style={dashed,draw=black,
      postaction={decorate},decoration={markings,mark=at position .55
        with {\arrow[draw=black]{>}}}}, 
    scalarbar/.style={dashed,draw=black,
      postaction={decorate},decoration={markings,mark=at position .55
        with {\arrow[draw=black]{<}}}}, 
    scalarnoarrow/.style={dashed,draw=black},
    electron/.style={draw=black,
      postaction={decorate},decoration={markings,mark=at position .55
        with {\arrow[draw=black]{>}}}}, 
    bigvector/.style={decorate, decoration={snake,amplitude=4pt}, draw},
}

\newcommand{\bea}{\begin{eqnarray}}
\newcommand{\eea}{\end{eqnarray}}
\newcommand{\be}{\begin{equation}}
\newcommand{\ee}{\end{equation}}
\newcommand{\mw}   {\mbox{$m_{W}$}}
\newcommand{\mwsq}   {\mbox{$m_{W}^2$}}

\newcommand{\mz}   {\mbox{$m_{Z}$}}
\newcommand{\mzsq}   {\mbox{$m_{Z}^2$}}

\newcommand{\muf}{\mu_F}

\title{NNLO mixed EW-QCD corrections to single vector boson production}

\ShortTitle{NNLO mixed EW-QCD corrections to single vector boson production}

\author{{Roberto Bonciani}\\
        Universit\`a  di  Roma  ``La  Sapienza''  and  INFN  Sezione  di  Roma1\\
        E-mail: \email{roberto.bonciani@roma1.infn.it}}

\author{{Federico Buccioni}\\
        Rudolf  Peierls  Centre  for  Theoretical  Physics,  Clarendon  Laboratory,  Parks  Road,  Oxford  OX1  3PU,  UK\\
        E-mail: \email{federico.buccioni@physics.ox.ac.uk}}
        
\author{\speaker{Narayan Rana}\\
        INFN  Sezione  di  Milano,  Via  Celoria  16,  20133  Milano,  Italy\\
        E-mail: \email{narayan.rana@mi.infn.it}}
        
\author{{Alessandro Vicini}\\
        Dipartimento  di  Fisica  ``Aldo  Pontremoli'',  University  of  Milano and INFN  Sezione  di  Milano,  Via  Celoria  16,  20133  Milano,  Italy\\
        E-mail: \email{alessandro.vicini@mi.infn.it}}
        
\abstract{We outline the computational details to obtain mixed EW-QCD corrections to on-shell production 
of a single vector boson at the LHC at two-loop level. 
We use the method of differential equation to obtain the pure virtual, 
real-virtual and double-real master integrals.
Finally, we obtain the ${\cal O}(\alpha \alpha_s)$ corrections to the total partonic cross section of the
process $q \bar{q} \rightarrow Z + X$.
}

\FullConference{14th International Symposium on Radiative Corrections (RADCOR2019)\\ 
9-13 September 2019\\
Palais des Papes, Avignon, France}

\begin{document}

\section{Introduction}
\noindent
%%%%%%%%%%%%%%%%%%%%%%%%%%%%%%%%%%%%%%%%%%%
The Drell-Yan (DY) production of a pair of high transverse momentum leptons
through the decay of a produced electrically neutral gauge boson ($\gamma^*/Z$) at hadron colliders,
is one of the most important processes for our understanding of Quantum Chromodynamics (QCD).
Due to its clean signature and abundant production at the hadron colliders,
it is used for the setting of several high-precision measurements of the electroweak (EW) sector
of the Standard Model (SM), for instance, a precise measurement of the weak mixing angle and
of the properties of the $Z$ boson.
Because of its high importance, precise theoretical predictions for 
the $Z$ boson DY production has always gained much attention. 
Following the pioneering calculations of the next-to-leading order (NLO) \cite{Altarelli:1979ub}
and next-to-next-to-leading order (NNLO) \cite{Hamberg:1990np,Harlander:2002wh} QCD corrections
to the total inclusive cross section, the fully differential description of the leptonic final state has been obtained in
Refs.~\cite{Anastasiou:2003ds,Melnikov:2006kv,Catani:2009sm,Gavin:2010az}.
Finally, the next-to-next-to-next-to-leading order (N$^3$LO)
QCD corrections have been presented in the threshold limit,
for both the total inclusive cross section \cite{Ahmed:2014cla,Li:2014bfa,Catani:2014uta}
and rapidity distribution \cite{Ahmed:2014uya,Lustermans:2019cau}.
While the QCD corrections have reached high precision, 
the NLO EW corrections,
as shown in Refs.~\cite{Baur:2001ze,CarloniCalame:2007cd,Arbuzov:2007db,Dittmaier:2009cr},
contribute at the ${\cal O}(1\%)$ level as far as the total cross section is concerned
and are comparable to that of the NNLO QCD contributions.
In Refs.~\cite{Berends:1987ab,Blumlein:2019srk}, NNLO Quantum Electrodynamics corrections 
have been presented which also turn out to be important.
Additionally in specific phase-space regions, 
kinematic distributions may be boosted, yielding corrections at the ${\cal O}(10\%)$ level or more.
Since control over the kinematic distributions in some cases is required at the per mille level
for the high-precision determination of EW parameters
(cf. Refs.~\cite{Alioli:2016fum,CarloniCalame:2016ouw,Bozzi:2015hha}
for specific cases),
the computation of the mixed QCD-EW corrections is quite demanding,
for both the study of the gauge boson resonances and of the high mass/momentum tails
of the kinematic distributions \cite{Balossini:2008cs,Balossini:2009sa}.
First steps have been taken by obtaining analytic expressions in
Refs.~\cite{Kilgore:2011pa,Dittmaier:2014qza,Dittmaier:2014koa,Dittmaier:2015rxo,Dittmaier:2016egk}.
They are then compared with the approximations available via Monte Carlo simulation
tools \cite{Barze:2012tt,Barze:2013fru}.
In these tools, the bulk of the leading effects, separately due to QCD and QED corrections,
can be correctly evaluated for several observables,
however, the remaining sub-leading QED effects and the genuine QCD-weak corrections are still missing,
Additionally, the matching of factorizable QCD and EW contributions depends on the chosen recipes
and hence introduces ambiguities to the estimation of theoretical uncertainties, which must be addressed.
% 
% Furthermore, a realistic estimate of the theoretical uncertainties must account for several sources
% of ambiguity related to the recipes used in the matching of separate results
% for the QCD and EW contributions to the scattering amplitude.
For these reasons the evaluation of the complete ${\cal O}(\alpha\alpha_s)$ corrections
to the DY processes is very much desirable.
In Refs.~\cite{deFlorian:2018wcj,Cieri:2018sfk,Delto:2019ewv} the mixed QCD-QED corrections
to the total inclusive cross section and transverse momentum spectrum of an on-shell $Z$ boson
have been obtained.
On another note, the Master Integrals (MIs) required to compute the complete
QCD-EW mixed corrections to DY process 
has been presented in Refs.~\cite{Bonciani:2016ypc,Heller:2019gkq}.

In this proceedings, we discuss the results, presented in Ref.~\cite{Bonciani:2019nuy}, for the total inclusive cross section
of production of an on-shell $Z$ boson in the quark-antiquark partonic channel,
including the complete set of QCD-EW corrections of ${\cal O}(\alpha\alpha_s)$.
% We retain the dependence on the massive states exchanged in the loops.
% As a consequence of that, the calculation involves a set of two-loop phase-space integrals,
% previously not available in the literature.
% Their analytic expression will be presented in a forthcoming paper. 
% %
% We also have the occasion to check the infrared structure of the corrections up to NNLO level,
% including the cases where a massive EW boson is exchanged.
% We verify the absence of initial state mass singularities proportional to a weak massive virtual
% correction to the quark-gluon splitting.
% 
% The calculation we are presenting in this letter is an important step towards
% the evaluation of the full set of QCD-EW corrections to the hadronic cross section. 

%%%%%%%%%%%%%%%%%%%%%%%%%%%%%%%%%%%%%%%%%%%
% 

\section{Theoretical framework}
%%%
\noindent
The total inclusive production cross section $\sigma_{tot}$ of a $Z$ boson at hadron colliders $(pp\to Z+X)$ can be written as 
\begin{align}
&\sigma_{tot} (\tau) = \sum_{i,j\in q,\bar q, g, \gamma}
\int {\rm d}x_1 {\rm d}x_2 \hat{f}_i (x_1) \hat{f}_j (x_2) \hat\sigma_{ij} (z) \,,
\label{eq:sigmatot-bare}
\end{align}
where $z=\frac{m_Z^2}{\hat{s}}$ and $\tau=\frac{m_Z^2}{S}$ are the ratio of the squared $Z$ boson mass, $\mz$, 
with $\hat{s}$ and $S$, the partonic and hadronic center of mass energy squared, respectively. 
$\hat{s}$ and $S$ are related by $\hat{s}=x_1 x_2 S$ through the Bjorken momentum fractions $x_1, x_2$.
$\hat{f}_i (x)$ is the bare parton density of the $i$-th incoming parton and 
$\hat\sigma_{ij}$ is the bare cross section of the partonic process  $ij\to Z+X$.
The sum over $i,j$ includes quarks ($q$), antiquarks ($\bar{q}$), gluons ($g$) and photons ($\gamma$). 
In the SM, we have the following double expansion of the partonic cross section in the electromagnetic and strong coupling constants, $\alpha$ and $\alpha_s$, respectively:
\begin{equation} 
\hat\sigma_{ij} (z) = \sum_{m,n=0}^{\infty}\alpha_s^m \alpha^n ~ \hat\sigma_{ij}^{(m,n)} (z) \, ,
\label{eq:sigmaexp}
\end{equation}
where $\hat\sigma_{ij}^{(m,n)}$ is the correction of ${\cal O}(\alpha_s^m\alpha^n)$ to the lowest-order inclusive total cross section $\hat\sigma_{ij}^{(0,0)}$ of the partonic scattering $ij \to Z$. 
% For a given initial state, the inclusive total cross section receives contributions from processes with different final state multiplicities, 
% due to real parton emissions. 
We consider the $q \bar{q}$ initiated scattering, specifically the case of an up-type quark: $q \bar{q} = u \bar{u}$.
The following scattering processes contribute to 
the complete set of ${\cal O}(\alpha\alpha_s)$ corrections to $\hat\sigma_{u\bar{u}}$:
\bea
&&u\bar u \to Z \, , \label{eq:proc1}\\
&&u\bar u \to Zg \, , \label{eq:proc2}\\
&&u\bar u \to Z\gamma \, , \label{eq:proc3}\\
&&u\bar u \to Zg\gamma \, , \label{eq:proc4}\\
&&u\bar u \to Z u\bar u \, , \label{eq:proc5}\\
&&u\bar u \to Z d\bar d \, , \label{eq:proc6}
\eea
where $d$ is a down-type massless quark. 
Results for the process (\ref{eq:proc4}) and QCD-QED contributions to process (\ref{eq:proc1})
have been presented in Ref.~\cite{Bonciani:2016wya} and Ref.~\cite{Kilgore:2011pa,H:2019nsw}, respectively.
The corresponding results for $d\bar d$ initiated subprocesses can be obtained from our results 
with proper replacements of the electric charge $(Q_f)$ and the third component of the weak isospin $(I_f^{(3)})$, for a fermion $f$.
% 
% with the replacements $Q_u\leftrightarrow Q_d$, $I_u^{(3)}\leftrightarrow I_d^{(3)}$, where $Q_f,\,I_f^{(3)}$ are the electric charge and the third component of the weak isospin, for a fermion $f$, respectively. 

The process (\ref{eq:proc1}) receives contributions (as shown in Fig.~\ref{fig:virt})
from two-loop $2 \to 1$ Feynman diagrams interfered with the Born process (the double-virtual contributions),
% tree-level $u \bar{u} \to Z$,
and interference of one-loop $2 \to 1$ Feynman diagrams (the virtual-virtual contributions).
\vspace{0.3cm}

\begin{minipage}[b]{0.97\textwidth}
\centering
\begin{tikzpicture}[scale=0.5]
 \draw[fermion] (-3.0,1.5) -- (0,0);
 \draw[fermion] (0,0) -- (-3.0,-1.5);
 \draw[photon] (-2.0,1.0) -- (-2.0,-1.0);
 \draw[gluon] (-1.0,0.5) -- (-1.0,-0.5);
 \draw[zboson] (0,0) -- (0.7,0);
 \draw[zboson] (1.8,0) -- (2.5,0);
 \draw[fermion] (3.5,-1.5) -- (2.5,0);
 \draw[fermion] (2.5,0) -- (3.5,1.5); 
 \node at (1.25,0) {$\times$};
\end{tikzpicture}
% \captionsetup{font=footnotesize}
% \captionof{figure}{Double-virtual contribution}
% \label{fig:dv}
% \end{minipage}
% \begin{minipage}[b]{0.49\textwidth}
%  \centering
\quad \quad
\begin{tikzpicture}[scale=0.5]
 \draw[fermion] (-2.0,1.5) -- (0,0);
 \draw[fermion] (0,0) -- (-2.0,-1.5);
 \draw[gluon] (-1.2,0.81) -- (-1.2,-0.83);
 \draw[zboson] (0,0) -- (0.7,0);
 \draw[zboson] (1.8,0) -- (2.5,0);
 \draw[photon] (3.7,0.81) -- (3.7,-0.81);
 \draw[fermion] (4.5,-1.5) -- (2.5,0);
 \draw[fermion] (2.5,0) -- (4.5,1.5); 
 \node at (1.25,0) {$\times$};
\end{tikzpicture}
\captionsetup{font=footnotesize}

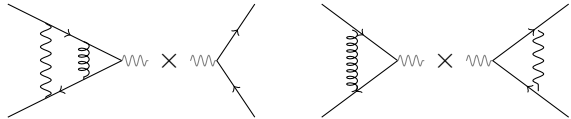
\captionof{figure}{Schematic diagrams for double-virtual and virtual-virtual contributions.}
\label{fig:virt}
\end{minipage}

The processes (\ref{eq:proc2})--(\ref{eq:proc3}) receive contributions
from one-loop $2 \to 2$ Feynman diagrams that have to be interfered
with the corresponding tree-level (as schematically presented in Fig.~\ref{fig:rv}). 
We refer them as the real-virtual contributions.
\vspace{0.3cm}

\begin{minipage}[b]{0.97\textwidth}
 \centering
\begin{tikzpicture}[scale=0.5]
 \draw[fermion] (-2.0,1.3) -- (0,1.3);
 \draw[fermion] (0,1.3) -- (0,-1.3);
 \draw[fermion] (0,-1.3) -- (-2.0,-1.3);
 \draw[gluon] (-1.0,1.3) -- (-1.0,-1.3);
 \draw[photon] (0.0,1.3) -- (0.7,1.3);
 \draw[zboson] (0,-1.3) -- (0.7,-1.3);
 \node at (1.25,1.3) {$\times$};
 \node at (1.25,-1.3) {$\times$}; 
 \draw[zboson] (1.7,-1.3) -- (2.5,-1.3);
 \draw[photon] (1.7,1.3) -- (2.5,1.3); 
 \draw[fermion] (2.5,1.3) -- (3.5,1.3);
 \draw[fermion] (2.5,-1.3) -- (2.5,1.3);
 \draw[fermion] (3.5,-1.3) -- (2.5,-1.3);
\end{tikzpicture}
\quad \quad
\begin{tikzpicture}[scale=0.5]
 \draw[fermion] (-2.0,1.3) -- (0,1.3);
 \draw[fermion] (0,1.3) -- (0,-1.3);
 \draw[fermion] (0,-1.3) -- (-2.0,-1.3);
 \draw[photon] (-1.0,1.3) -- (-1.0,-1.3);
 \draw[gluon] (0.0,1.3) -- (0.7,1.3);
 \draw[zboson] (0,-1.3) -- (0.7,-1.3);
 \node at (1.25,1.3) {$\times$};
 \node at (1.25,-1.3) {$\times$}; 
 \draw[zboson] (1.7,-1.3) -- (2.5,-1.3);
 \draw[gluon] (1.7,1.3) -- (2.5,1.3); 
 \draw[fermion] (2.5,1.3) -- (3.5,1.3);
 \draw[fermion] (2.5,-1.3) -- (2.5,1.3);
 \draw[fermion] (3.5,-1.3) -- (2.5,-1.3);
\end{tikzpicture}
\quad \quad
\begin{tikzpicture}[scale=0.5]
 \draw[fermion] (-2.0,1.3) -- (0,1.3);
 \draw[fermion] (0,1.3) -- (0,0);
 \draw[fermion] (0,0) -- (-1,-1.3); 
 \draw[fermion] (-1.0,-1.3) -- (-2.0,-1.3);
 \draw[photon] (0,0) -- (0,-1.3);
 \draw[photon] (-1.0,-1.3) -- (0,-1.3);
 \draw[gluon] (0.0,1.3) -- (0.7,1.3);
 \draw[zboson] (0,-1.3) -- (0.7,-1.3);
 \node at (1.25,1.3) {$\times$};
 \node at (1.25,-1.3) {$\times$}; 
 \draw[zboson] (1.7,-1.3) -- (2.5,-1.3);
 \draw[gluon] (1.7,1.3) -- (2.5,1.3); 
 \draw[fermion] (2.5,1.3) -- (3.5,1.3);
 \draw[fermion] (2.5,-1.3) -- (2.5,1.3);
 \draw[fermion] (3.5,-1.3) -- (2.5,-1.3);
\end{tikzpicture}
\captionsetup{font=footnotesize}

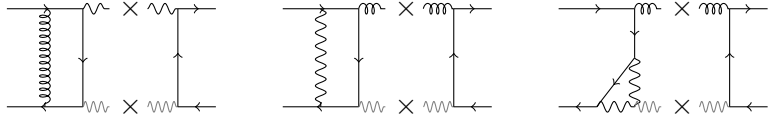
\captionof{figure}{Schematic diagrams for the real-virtual contribution.}
\label{fig:rv}
\end{minipage}

The last three processes, (\ref{eq:proc4})--(\ref{eq:proc6}),
receive contributions from tree-level $2 \to 3$ Feynman diagrams interfered
with themselves (the double-real contributions as shown in Fig.~\ref{fig:dr}).
% and we refer to them as double-real corrections.
\vspace{0.3cm}

\begin{minipage}{0.97\textwidth}
\centering
\begin{tikzpicture}[scale=0.5]
 \draw[fermion] (-1.0,1.3) -- (0,1.3);
 \draw[fermion] (0,1.3) -- (0,-1.3);
 \draw[fermion] (0,-1.3) -- (-1.0,-1.3);
 \draw[gluon] (0.0,0.0) -- (0.7,0);
 \draw[photon] (0.0,1.3) -- (0.7,1.3);
 \draw[zboson] (0,-1.3) -- (0.7,-1.3);
 \node at (1.25,1.3) {$\times$};
 \node at (1.25,0) {$\times$}; 
 \node at (1.25,-1.3) {$\times$}; 
 \draw[zboson] (1.7,-1.3) -- (2.5,-1.3);
 \draw[photon] (1.7,1.3) -- (2.5,1.3); 
 \draw[gluon] (1.7,0) -- (2.5,0); 
 \draw[fermion] (2.5,1.3) -- (3.5,1.3);
 \draw[fermion] (2.5,-1.3) -- (2.5,1.3);
 \draw[fermion] (3.5,-1.3) -- (2.5,-1.3);
\end{tikzpicture}
\quad \quad
\begin{tikzpicture}[scale=0.5]
 \draw[fermion] (-3.5,1.3) -- (-2.5,0);
 \draw[fermion] (-2.5,0) -- (-3.5,-1.3);
 \draw[gluon] (-2.5,0) -- (-1.5,0);
 \draw[fermion] (-1.5,0) -- (-0.5,1.3);
 \draw[fermion] (-0.5,-1.3) -- (-1.5,0); 
 \draw[zboson] (-1.1,0.5) -- (-0.3,-0.0); 
 \draw[zboson] (0.5,-0.4) -- (2.0,-1.3); 
 \node at (0, 1.3) {$\times$}; 
 \node at (0,-0.2) {$\times$}; 
 \node at (0,-1.3) {$\times$}; 
 \draw[fermion] (0.5,1.3) -- (3.5,1.3);
 \draw[fermion] (3.5,-1.3) -- (0.5,-1.3);
 \draw[photon] (2.5,1.3) -- (2.5,-1.3);
\end{tikzpicture}
\quad \quad
\begin{tikzpicture}[scale=0.5]
 \draw[fermion] (-3.5,1.3) -- (-2.5,0);
 \draw[fermion] (-2.5,0) -- (-3.5,-1.3);
 \draw[gluon] (-2.5,0) -- (-1.5,0);
 \draw[fermion] (-1.5,0) -- (-0.5,1.3);
 \draw[fermion] (-0.5,-1.3) -- (-1.5,0); 
 \draw[zboson] (-1.1,0.5) -- (-0.3,-0.0); 
 \draw[zboson] (0.5,0.0) -- (2.0,0.0); 
 \node at (0, 1.3) {$\times$}; 
 \node at (0,-0) {$\times$}; 
 \node at (0,-1.3) {$\times$}; 
 \draw[fermion] (0.5,1.3) -- (3.5,1.3);
 \draw[fermion] (3.5,-1.3) -- (0.5,-1.3);
 \draw[photon] (2.0,1.3) -- (2.0,-1.3);
\end{tikzpicture}
\captionsetup{font=footnotesize}

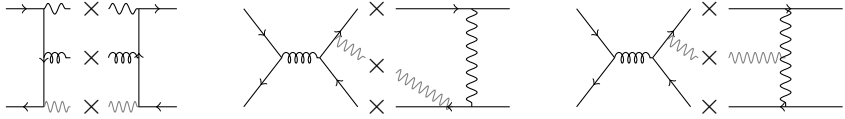
\captionof{figure}{Schematic diagrams for the double-real contributions}
\label{fig:dr}
\end{minipage}
\vspace{0.3cm}

The complete ${\cal O}(\alpha\alpha_s)$ corrections can be organized
in two subsets, both gauge invariant: QCD-QED and QCD-weak contributions.
Processes (\ref{eq:proc1})--(\ref{eq:proc5}) contribute to the former,
and processes (\ref{eq:proc1}), (\ref{eq:proc2}), (\ref{eq:proc5}) and (\ref{eq:proc6}) to the latter.
On the other hand, depending
on the presence of one real or virtual photon, of one virtual $Z$ boson,
or of one/two virtual $W$ bosons, 
we can organize the full corrections in three groups.
We note that the last two groups are separately gauge invariant.
Furthermore, we do not include the processes
with the emission of one extra massive on-shell gauge boson, as their measurement
depends on the details of the experimental event selection.

% The amplitude of the two tree-level processes (\ref{eq:proc5}) and (\ref{eq:proc6})
% has two components of ${\cal O}(\sqrt{\alpha} \alpha_s)$ (an internal gluon exchange)
% and ${\cal O}( \sqrt{\alpha} \alpha )$ (an internal weak boson exchange),
% respectively and their interference is, therefore, of ${\cal O}(\alpha^2 \alpha_s)$. 

%%%%%%%%%%%%%%%%%%%%%%%%%%%%%%%%%%%%%%%%%%%
\section{Computational details}

We follow the diagrammatic approach to obtain all the relevant contributions.
We treat all the processes, virtual and real emission contributions, with the same algorithmic approach.
Two independent parallel computations have been performed to have consistent check on our results.
In one process, we obtain all the Feynman diagrams contributing to a given amplitude with 
{\texttt{QGRAF}} \cite{Nogueira:1991ex} and in another we use {\texttt{FeynArts}} \cite{Hahn:2000kx}.
Next, we perform algebraic simplifications using in-house {\texttt{FORM}} \cite{Vermaseren:2000nd}
and {\texttt{Mathematica}} routines.
% Here we note that the presence of $\gamma_5$ in the computation makes it complicated.
The scalar Feynman integrals are reduced to MIs using 
integration-by-parts (IBP) \cite{Tkachov:1981wb,Chetyrkin:1981qh,Laporta:2001dd} and Lorentz-invariance (LI) identities \cite{Gehrmann:1999as}. 
The reduction procedure is performed using
\texttt{LiteRed} \cite{Lee:2012cn,Lee:2013mka} in the first procedure
and using
\texttt{Kira} \cite{Maierhoefer:2017hyi} and
\texttt{Reduze 2} \cite{Studerus:2009ye,vonManteuffel:2012np} in the second.
The entire computation is carried out within dimensional regularization in $D=4-2 \varepsilon$ space-time dimensions.
%
%%%%%%%%%%%%%%%%%%%%%%%%%%%%%%%%%%%%%%%%%%%
% \subsection{Reverse Unitarity and the Master Integrals}
% 
Then, we use the method of differential equations
\cite{Kotikov:1990kg,Remiddi:1997ny,Gehrmann:1999as,Argeri:2007up,Henn:2014qga,Ablinger:2015tua,Ablinger:2018zwz}
to obtain the MIs, for both the pure virtual and real emission corrections.
In the latter case, the phase-space delta functions are dealt using the reverse unitarity technique
\cite{Anastasiou:2002yz,Anastasiou:2012kq}, as follows
% which is based on the observation that the replacement known as Cutkowsky rule
% holds in terms of distributions: 
% 
\begin{equation}
\delta(p^2-m^2) \rightarrow 
\frac{1}{2\pi i} \left( \frac{1}{p^2-m^2+i\eta} - \frac{1}{p^2-m^2-i\eta} \right). 
\end{equation}
Thus we transform the integration over the full phase space of the additional parton/s for processes
(\ref{eq:proc2})--(\ref{eq:proc6}), into the evaluation of the two-loop integrals with cut propagators.
Apart from imposing an on-shell condition on the lines that correspond to the final-state particles,
the integrals behave as loop integrals and hence one can use the techniques like IBP and differential equations
to solve them. Below, we present a short description of the method (See \cite{Ablinger:2018zwz,Blumlein:2018cms,Ablinger:2017hst} for details).

% Let's consider a set of $n$ MIs ${\cal I} = (I_1,\ldots,I_n)$ within the same topology. 
The MIs are functions of the space--time dimension $D$ and the variable $z$.
The basic idea is to obtain a set of differential equations of the MIs by performing differentiation \textit{w.r.t} 
$z$ and then to use the IBP identities.
In all the cases at hand, we obtain a system of differential equations which can be organized in a block-triangular form,
with most of the blocks being $1\times1$ and the rest being $2\times2$. 
For the coupled sub-systems, we obtain a second order linear differential equation by uncoupling and solve them using
the method of variation of constant.
The solution for each integral is obtained in series expansion in $\varepsilon$ up to the required order.
In calculating the MIs,
the package {\texttt{HarmonicSums}} \cite{Ablinger:2010kw,Ablinger:2013cf,Ablinger:2014rba} has been used.

The pure virtual MIs are presented in 
\cite{Aglietti:2003yc,Aglietti:2004tq,Aglietti:2004ki,Aglietti:2007as,Bonciani:2010ms,Kotikov:2007vr},
considering an off-shell $Z$ boson.
In case of a single internal massive line, 
the solutions for these integrals contain generalised harmonic polylogarithms (GPLs)
or harmonic polylogarithms (HPLs) \cite{Goncharov:polylog,Goncharov2007,Remiddi:1999ew,Vollinga:2004sn} 
over the alphabet
\begin{equation}
 \Big\{ \frac{1}{z}, \frac{1}{1-z}, \frac{1}{1+z} \Big\}\,.
\end{equation}
In case of two internal massive lines, square-root letters appear. To rationalize them, we introduce the Landau variable $x$ as
\begin{equation}
 \frac{\hat s}{m_Z^2} = - \frac{(1-x)^2}{x} \,.
\end{equation}
and the new letters contributing to the alphabet are
\begin{equation}
 \Big\{ \frac{1}{1-x+x^2}, \frac{x}{1-x+x^2} \Big\}\,,
\end{equation}
which are the cyclotomic extension \cite{Ablinger:2011te} of the HPLs. 
Since in our case the $Z$ boson is on-shell,
we have obtained these integrals taking the appropriate on-shell limit,
\textit{i.e.} evaluating the GPLs or HPLs at $z=1$ or equivalent limit of $x$.
All the constants appearing, can be reduced to the basis introduced in Ref.~\cite{Henn:2015sem}.
The off-shell integrals and most of their on-shell limit have been checked using 
{\texttt{FIESTA}} \cite{Smirnov:2008py,Smirnov:2009pb,Smirnov:2015mct}.

The two- and three-body phase-space MIs with only gluon or photon lines are already
available in the literature \cite{Anastasiou:2012kq}.
To validate our routines developed for the present calculation, however,
we have recomputed them and found complete agreement with the known expressions.
We have computed all the new MIs, with one or two internal massive lines.
% with the method of differential equations. 
During the computation, we need to introduce several other variables \textit{e.g.} $w$ and $\rho$ as
\begin{equation}
 z = \frac{w}{(1+w)^2}  \,; \quad z = \frac{\rho}{1-\rho+\rho^2}
\end{equation}
to obtain
a rationalized alphabet with addition of the following letters
\begin{equation}
 \Big\{ \frac{1}{1+x+x^2}, \frac{x}{1+x+x^2}, \frac{1}{1+x^2}, \frac{x}{1+x^2} \Big\}\,.
\end{equation}
The boundary conditions are obtained by explicitly calculating the MIs in the soft limit ($z\to 1$).

After the phase-space integration,
the various contributions to partonic total cross section depends solely on the variable $z$.
The virtual contributions are proportional to $\delta(1-z)$.
% The proportionality constants are found from the on-shell limit of the virtual MIs, 
% i.e. evaluating the corresponding generalised harmonic polylogarithms (GPLs)
% \cite{Goncharov:polylog,Goncharov2007,Remiddi:1999ew,Vollinga:2004sn} at $z=1$.
The part that corresponds to processes (\ref{eq:proc2})--(\ref{eq:proc6})
is expressed almost entirely in terms of $\delta(1-z)$ and of GPLs,
or cyclotomic HPLs \cite{Ablinger:2011te}, functions of $z$.
Three MIs appearing in processes (\ref{eq:proc5}) and (\ref{eq:proc6})
satisfy elliptic differential equations,
whose homogeneous behaviour has already been studied in Ref.~\cite{Aglietti:2007as}.
We have obtained their complete solution with a series expansion around $z=1$
(see for instance \cite{Pozzorini:2005ff,Aglietti:2007as,Blumlein:2017dxp,Lee:2017qql,Lee:2018ojn,Bonciani:2018uvv,Blumlein:2019oas}).
In the computation of the MIs, the mass of the $W$ boson is set equal to $\mz$, the mass of $Z$ boson,
to avoid the presence of an additional energy scale in the problem,
which would make the analytical solution of the differential equations in terms of known functions
more complicated. However, to obtain precise results, we can perform an expansion of the integrand in powers of
the ratio  $\delta_m^2=(\mzsq-\mwsq)/\mzsq$,
and reduce all the terms of the series to a combination of the same basic equal-mass MIs.
% 
% While for the virtual corrections this choice is not strictly necessary,
% since the knowledge of the MIs for off-shell $Z$ would allow
% for a complete and exact calculation in the case of $m_W \not = m_Z$,
% the reduction of one mass scale in the computation of the real emission processes is in fact very effective
% and reduces the complication of the calculation.
% %
% Moreover, the equal-mass choice does not prevent us from obtaining an analytical solution
% with arbitrary precision for each of the affected MIs.
% In fact, we can perform an expansion of the integrand in powers of
% the ratio  $\delta_m^2=(\mzsq-\mwsq)/\mzsq$,
% and reduce all the terms of the series to a combination of the same basic equal-mass MIs.
%
We note that the couplings of the $Z$ boson to fermions are expressed in terms of the physical
value of the weak mixing angle $\sin^2\theta_W=1-\mwsq/\mzsq$.

%%%%%%%%%%%%%%%%%%%%%%%%%%%%%%%%%%%%%%%%%%%
\section{Ultraviolet and infrared singularities}

In general, we require the renormalization of the couplings and the fields up to ${\cal O}(\alpha\alpha_s)$.
However, the Born process does not contain $\alpha_s$, hence no ultraviolet (UV) renormalization is required for the same.
The $Z$ boson field and the EW couplings also do not receive ${\cal O}(\alpha_s)$ renormalization corrections.
On the other hand, the quark field receives EW $({\cal O}(\alpha))$ and mixed QCD-EW $({\cal O}(\alpha\alpha_s))$ 
renormalization corrections which we consider in the on-shell scheme.
To simplify the UV singularities arising from the EW sector, we perform the calculation in the EW background field gauge (BFG) \cite{Denner:1994xt}.
The advantage of using BFG is that, on the one hand, the combination of one-particle-irreducible vertex 
and external quark wave function corrections are UV finite,
and on the other, the external $Z$ boson wave function and the lowest-order coupling renormalization corrections, 
whose combination, order-by-order in perturbation theory, is also UV finite.
We remark that, in general notion, the EW gauge sector of the SM Lagrangian depends on three parameters $(g,g',v)$, 
the two gauge couplings and the Higgs-doublet vacuum expectation value. 
However, we choose to use the combination of $(G_\mu,\mw, \mz)$, respectively the Fermi constant, the $W$ and $Z$ boson masses,
after introducing counterterms and renormalized parameters. 
See Ref.~\cite{Degrassi:2003rw} for the description of the additional counterterms appearing for such replacement.
An alternative scheme with the effective leptonic weak mixing angle as input parameter
has been discussed in Ref.~\cite{Chiesa:2019nqb}.

% A subset of the EW corrections can be reabsorbed in a redefinition of the weak mixing angle that appears in the vector coupling of the $Z$ boson to fermions.
% These corrections are split, in the EW BFG, in two UV-finite groups, one due to vertex corrections, the other due to external $\gamma-Z$ corrections and to the weak mixing angle counterterm (a shortcut for a combination of $W$ and $Z$ mass counterterms). In BFG the second group vanishes, because of a Ward identity \cite{Denner:1994xt}
% satisfied by the $\gamma-Z$ wave-function correction.

%%%%%%%%%%%%%%%%%%%%%%%%%%%%%%%%%%%%%%%%%%%
% \subsection{Infrared singularities and mass factorization}

The infrared (IR) singularities arising in this scenario are of two types by nature: soft singularities due to soft massless bosons
and collinear singularities coming from collinear partons.
As the ${\cal O}(\alpha\alpha_s)$ corrections are organized in two gauge invariant subsets: QCD-QED and QCD-weak contributions,
we study the IR singularities for each subset.
For both the subsets, once all the degenerate states are summed up, \textit{i.e.} the processes (\ref{eq:proc1})--(\ref{eq:proc6}),
the soft and final-state collinear singularities cancel. What remains are the collinear singularities arising from initial states,
which are removed by mass factorization.
For the QCD-QED subset, a new type of mass factorization kernel $\Gamma_{ij}$ with mixed non-factorizing contributions, appears \cite{deFlorian:2015ujt}.
However initial state collinear singularities in the QCD-weak case are of QCD origin only.
% 
% The former involve the exchange of two massless bosons, yielding the maximal degree of infrared singularity at the second perturbative order, 
% i.e. $\varepsilon^{-4}$. The latter have only the poles due to a soft and/or collinear gluon.
% The cancellation of the soft singularities takes place separately in the two subsets, once the contribution of virtual corrections 
% and of the corresponding soft real emissions are combined. To be more precise, for the QCD-QED subset, the process (\ref{eq:proc5}) 
% does not yield soft singularities, so that the cancellation takes place when the processes (\ref{eq:proc1})--(\ref{eq:proc4}) are combined. 
% In the case of the QCD-weak subset, soft singularities appear only in processes (\ref{eq:proc1}) and (\ref{eq:proc2}) and cancel 
% when the two are summed. When we consider the combination of the cross sections of the processes (\ref{eq:proc1})--(\ref{eq:proc6}) 
% we are thus left with initial state collinear singularities only. The processes (\ref{eq:proc1})--(\ref{eq:proc5}) 
% contribute to initial-state collinear singularities within the QCD-QED subset, 
% while in the QCD-weak case only processes (\ref{eq:proc1})--(\ref{eq:proc2}) have initial state collinear singularities of QCD origin. 
% These singularities can be removed by mass factorization. 
The mass factorization introduces the physical parton densities $f_i(x,\muf)$, at the factorization scale $\muf$, 
which are defined through the mass factorization kernels $\Gamma_{ij}$ as follows
\begin{equation}
\hat{f}_i = f_j \otimes \Gamma_{ij} \,.
\label{eq:baretophysicalPDF}
\end{equation}
The kernels admit a series expansion in $\alpha$ and $\alpha_s$.
% \begin{equation}
% \Gamma_{ij}
% =
% \sum_{m,n=0}^{\infty}\alpha_s^m \alpha^n \Gamma_{ij}^{(m,n)}\,,
% \label{eq:gammaexp}
% \end{equation}
% 
% where $\Gamma_{ij}^{(1,0)}$ is the QCD leading order (LO) splitting kernel, 
% $\Gamma_{ij}^{(0,1)}$ is its QED analogue and $\Gamma_{ij}^{(1,1)}$ is the mixed QCD-QED contribution to the splitting kernels, 
% computed in Ref.~\cite{deFlorian:2015ujt}. 
%
Finally, using Eq.~(\ref{eq:baretophysicalPDF}) in Eq.~(\ref{eq:sigmatot-bare}), 
we obtain the total cross section expressed in terms of subtracted, finite, partonic cross sections $\sigma_{ij}(z, \muf)$:
\be
\sigma_{tot}(z) =\!\!\!\!\!\!
\sum_{i,j\in q,\bar q, g, \gamma}
\!\!\!\!\!\!
\int {\rm d}x_1 {\rm d}x_2  
f_i(x_1,\muf) f_j(x_2,\muf) \sigma_{ij} (z, \muf)\, .
\ee
The $\sigma_{ij}$ also admits a perturbative expansion in powers of $\alpha$ and $\alpha_s$, in analogy to Eq.~(\ref{eq:sigmaexp}). 
% In this letter, we present the results for $\sigma_{u\bar{u}}^{(1,1)}$.
% 
% In processes (\ref{eq:proc1}) and (\ref{eq:proc2}) the weak virtual correction to the splitting vertex $q\to qg$
% might induce an additional contribution to the subtraction kernel $\Gamma_{ij}^{(1,1)}$. 
% However, we have checked that such a term vanishes, in the massless quark case, as a consequence of the conservation of the vector and axial-vector currents.

\section{Results}

We obtain the results for $\sigma_{u\bar{u}}^{(1,1)}$. As mentioned earlier, $\sigma_{u\bar{u}}^{(1,1)}$ depends on $z$
through $\delta(1-z)$ and GPLs \cite{Goncharov:polylog,Goncharov2007,Remiddi:1999ew,Vollinga:2004sn} 
or cyclotomic HPLs \cite{Ablinger:2011te}. Additionally, the contributions from elliptic integrals appear as series 
expansion around threshold \textit{i.e.} $z\rightarrow1$.
With the $\delta(1-z)$ contribution, apart from the multiple zeta values \cite{Blumlein:2009cf}, 
the following constants appear
\begin{equation}
\{ \ln 2, {\rm Li}_4 (1/2), {\rm GI}[r_2], {\rm GI}[0, r_2], {\rm GI}[0, 1, r_4] \}\,,
\end{equation}
where $r_i$ are the sixth root of unity defined as
\begin{equation}
 r_1 = e^{i \pi/3}\,,~~ r_2 = e^{-i \pi/3}\,,~~ r_3 = e^{2 i \pi/3}\,,~~ r_4 = e^{- 2 i \pi/3}\,.
\end{equation}
Additionally, the computation of the boundary constants of the differential equations
for the double-real emissions, generates the following cyclotomic constants \cite{Ablinger:2011te,Broadhurst:1998rz}
\begin{align}
& H[\{4, 1\}, 0, -1, (-1)^{2/3}], ~ 
  H[\{4, 1\}, 0, 0, (-1)^{2/3}], ~
  H[\{4, 1\}, \{3, 0\}, -1, (-1)^{2/3}], ~
\nonumber\\
& H[\{4, 1\}, \{3, 1\}, -1, (-1)^{2/3}], ~
  H[\{4, 1\}, \{3, 0\}, 0, (-1)^{2/3}], ~ 
  H[\{4, 1\}, \{3, 1\}, 0, (-1)^{2/3}]\,.
\end{align}
Here, the letters of cyclotomy 3 and 4 are
\begin{align}
 f_{\{3,0\}} (x) = \frac{1}{1+x+x^2}\,,
 f_{\{3,1\}} (x) = \frac{x}{1+x+x^2}\,,
 f_{\{4,0\}} (x) = \frac{1}{1+x^2}\,,
 f_{\{4,1\}} (x) = \frac{x}{1+x^2}\,.
\end{align}
In order to anticipate the relative size of the different sets of corrections,
we define:
\bea
\alpha_s\alpha \sigma_{u\bar{u}}^{(1,1)}
\!\!\!\!&=&\!\!
\sigma_{u\bar{u}}^{(0)}
\left(
\Delta_{u\bar{u},\gamma}^{(1,1)}
+
\Delta_{u\bar{u},Z}^{(1,1)}
+
\Delta_{u\bar{u},W}^{(1,1)}
\right)\,\,\,\,\,\,\,
\eea
where $\sigma_{u\bar{u}}^{(0,0)} \equiv \sigma_{u\bar{u}}^{(0)} \delta(1-z) =
4\sqrt{2}  G_\mu (\pi/N_c) (C_{v,u}^2+C_{a,u}^2) \delta(1-z)$
is the Born cross section.
% of the process $u\bar u\to Z$, with 
$N_c$ is the number of colours
and $C_{v/a,u}$ are the vector/axial-vector couplings of the $Z$ boson to the up quark.
$\Delta_{u\bar{u},K}^{(1,1)}$ with $K=\gamma,Z,W$ are the corrections due to the exchange
of a photon, a $Z$ boson, and of one or two $W$ boson/s including the lowest order charge renormalization counterterms,
respectively.
We introduce the NLO-QCD correction to the same partonic process,
defined as $\alpha_s \sigma_{u\bar{u}}^{(1,0)} = \sigma_{u\bar{u}}^{(0)} \,\, \Delta_{u\bar{u}}^{(1,0)}$,
to have a comparison.
\begin{figure}[ht]
\centering
\includegraphics[width=0.47\textwidth]{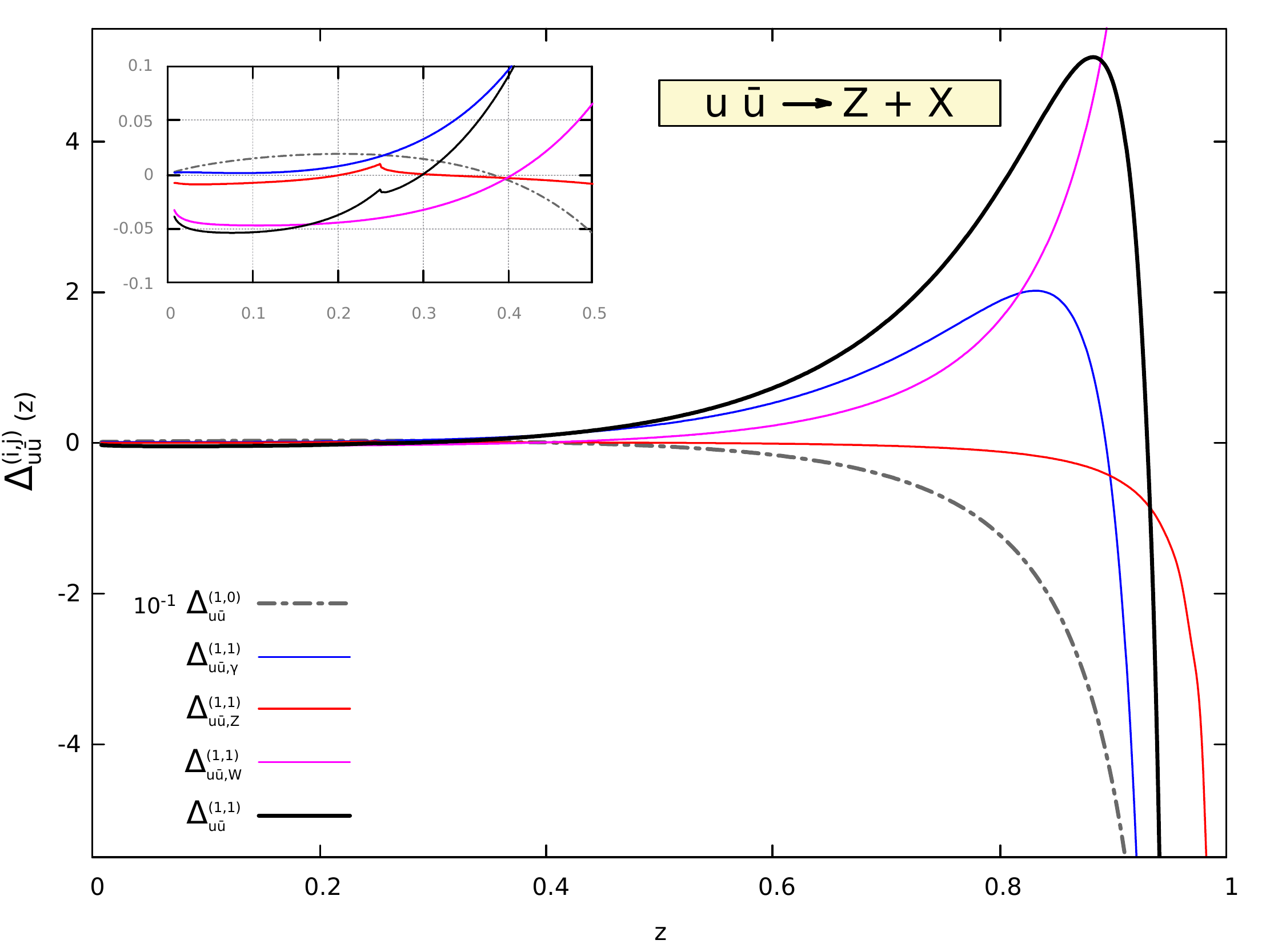}
  \caption{$\Delta_{u\bar{u}}^{(1,0)}$ in grey dashed, $\Delta_{u\bar{u},K}^{(1,1)}$ with $K=\gamma,Z,W$ in blue, red, and magenta, respectively,
    and their sum in black solid,
    as a function of the partonic variable $z$.
    $\Delta_{u\bar{u}}^{(1,0)}$
    is divided by a factor 10.
    \label{fig:deltas}}
\end{figure}
In Figure \ref{fig:deltas} we present
the contributions of the different subsets, $\Delta_{u\bar{u},K}^{(1,1)}$ with $K=\gamma,Z,W$, and their sum.
We also present $\Delta_{u\bar{u}}^{(1,0)}$, divided by a factor 10.
We exclude from the plot all the
contributions proportional to $\delta(1-z)$.
% while we keep all the plus-distribution terms, limiting the plot at $z=0.99$.
For the numerics, we use the following input parameters:
\begin{center}
\begin{tabular}{l l}
\hline\hline
 $m_W=80.385$ GeV & $m_Z=91.1876$ GeV \\
 $m_t=173.5$ GeV & $m_H=125$ GeV\\
 $G_\mu=1.1663781\times 10^{-5}$ GeV$^{-2}$ & $\alpha_s(\mz)=0.118$ \\
\hline\hline
\end{tabular}
\end{center}
% 
% $m_W=80.385$ GeV, $m_Z=91.1876$ GeV, $G_\mu=1.1663781\times 10^{-5}$ GeV$^{-2}$,
% $m_t=173.5$ GeV, $m_H=125$ GeV, $\alpha_s(\mz)=0.118$.
$m_t$ and $m_H$ are the top quark and Higgs boson mass, respectively.
We set the factorisation scale $\mu_F=\mz$.

% We observe that, in the high-energy limit ($z\to 0$), the cross sections
% are damped by the incoming flux factor, proportional to $z$.
% The divergent behaviour for $z\to 1$, due to the exchange of at least one massless boson,
% is also evident for all the contributions.
% The values of the EW charges, in the two subsets with one $Z$ (red) or with one/two $W$s exchange (magenta),
% are responsible for the different size and for the opposite sign of the two contributions,
% visible in the $z\to 1$ limit.
% We observe that in the case of the $d\bar d\to Z+X$ process,
% the contributions with one/two $W$s exchange have
% similar size but opposite sign.
% The total contribution to the hadron-level cross section
% from this subset of diagrams of the two partonic processes
% is expected to undergo an important cancellation,
% modulated by the convolution with the proton PDFs.
% The QCD-QED corrections, shown in blue in Figure \ref{fig:deltas}, are not monotonic,
% contrary to the NLO-QCD ones and have a maximum for $z\sim 0.85$.
% They are smaller than the QCD-weak contribution for $z\in [0.8,0.9]$,
% but become larger in absolute size when $z\to 1$,
% because of the higher power of the threshold logarithms.
% The possibility of having a second $Z$ boson in a resonant configuration
% yields the kink of the $\Delta_{u\bar{u},Z}^{(1,1)}$ curve (red) at $z=1/4$,
% as it can be observed in the inset of Figure \ref{fig:deltas}.

\section{Conclusion}
In this proceeding, we discuss Ref.~\cite{Bonciani:2019nuy}, where
we have presented the first results for the total inclusive partonic cross section
for the process $q\bar q\to Z+X$, including the exact ${\cal O}(\alpha \alpha_s)$ corrections,
with both photon and $W/Z$ boson exchanges.
The results are analytic and are expressed in terms of GPLs,
but also contain three elliptic MIs, which have been computed with a series expansion around $z=1$.
The universal structure of the infrared singularities along with the final cancellation among all sub-processes
to produce a subtracted, finite, partonic cross section, provides also a strong check on our calculation.
The computation represents an important step towards the evaluation of the hadron-level cross section
for $Z$ production at this perturbative order.

\begin{acknowledgments}
  N.R. thanks the CERN Theory Department for hospitality and support during the completion of this work.
  A.V. is supported by the European Research Council
  under the European Unions Horizon 2020 research and innovation Programme (grant agreement number 740006).
F.B. warmly thanks the Physics Institute of the University
of Zurich where large part of this work was carried out and
acknowledges support from the 
Swiss National Science Foundation (SNF) under contract BSCGI0-157722. 
The research of F.B. was partially supported by the
ERC Starting Grant 804394 {\sc{hip}QCD}. 
R.B. and N.R. acknowledge the COST (European Cooperation in Science and
Technology) Action CA16201 PARTICLEFACE for partial support.
\end{acknowledgments}

\bibliographystyle{JHEP}
% \bibliography{mybib}

\providecommand{\href}[2]{#2}\begingroup\raggedright\endgroup

\end{document}